\documentclass[12pt]{iopart} 
\eqnobysec
\jl{1}

\begin{document}
\title{Vacuum charge fractionization re-examined }
\author{Y Nogami\footnote[1]{E-mail address: nogami@mcmaster.ca}}
\address{Department of Physics and Astronomy, McMaster
University,\\ Hamilton, Ontario, Canada L8S 4M1}  

\date{Draft of February 13, 2002}

\begin{abstract}
We consider a model of a quantized fermion field that is based on the 
Dirac equation in one dimensional space and re-examine how 
the fermion number of the vacuum, or the vacuum charge, 
varies when an external potential is switched on. With this 
model, fractionization of the vacuum charge has been 
illustrated in the literature by showing that the external 
potential can change the vacuum charge from zero to a 
fractional number. Charge conservation then appears 
violated in this process. This is because the charge that 
has been examined in this context is only a part of the 
total charge of the vacuum. The total charge is conserved. 
It is not fractionalized unless the Dirac equation has a 
zero mode. Two other confusing aspects are discussed. One 
is concerned with the usage of the continuum limit and the 
other with the regularization of the current operator. 
Implications of these aspects of the vacuum problem are 
explored. 
\end{abstract}

\pacs{PACS 03.65.-w, 11.10.-z, 12.39.Ba, 70.10.Ca}


\section{introduction}
The vacuum charge is the fermion number of the vacuum in
units of the charge of the particle that is associated 
with the field under consideration. Throughout this 
paper we use the words ``fermion number" and ``charge" 
interchangeably. Jackiw and Rebbi \cite{1} pointed out 
that, in the presence of a zero mode (a normalizable 
eigenstate of the Dirac equation with energy zero), the
vacuum charge of the associated quantum field can take 
half-integral values. Their observation was followed by 
a large number of papers, mostly in the late 1970s and 1980s, 
in which various such possibilities were explored \cite{2}. 
According to those analyses there are situations such
that the vacuum charge can take not only half-integral 
but any fractional, irrational values. The notion of 
fractionization of the vacuum charge has been discussed 
in many different areas of physics. 

The fractional charges that have been examined so far 
can be classified into two types, A and B, depending on 
how the charge varies when the external potential (or
background field) involved is continuously varied. Type A 
is such that the charge remains invariant under the the 
variation of the external potential. The half-integral 
charge that was found by Jackiw and Rebbi \cite{1} is of 
this type. There are other examples of this type in which 
the charge can take values $\pm \frac{1}{3}$ and 
$\pm \frac{2}{3}$ in certain units \cite{2a}. In type B, 
the charge (as it appeared in the literature) varies 
continuously as the external potential is varied. In type 
A it is essential that the external potential has a 
certain topological structure in terms of its large 
distance behavior. We do not question type A which we 
think is well understood. Rather we focus on type B. 
In type B it may appear as if some topology of the 
external potential is involved. As we will show in due 
course, however, the topology related to type B is 
actually a trivial one.  

In this paper we consider a fermion field which is based 
on the Dirac equation in one-dimensional space and 
re-examine the charge of its vacuum. This model is a 
typical one with which fractionization of the vacuum 
charge has been illustrated in the literature \cite{2}. 
We assume that there is no zero mode so that the charge 
fractionization of type A is absent. In the 
fractionization of type B, when an external potential 
is switched on, the vacuum charge changes from zero to 
a fractional value. It then appears that the charge is 
not conserved in this process. This is because, as we 
will emphasize, the charge that has been examined in 
this context in the literature is only a part of the 
total charge. Despite the long history of the vacuum 
problem, its aspect regarding fractionization versus 
conservation of the vacuum charge has not been fully 
addressed. The purpose of this paper is to clarify and 
explore implications of this and related aspects.   

Throughout this paper we follow Dirac's hole theory.
Alternatively one can use the second-quantized quantum 
field theory. We, however, find the hole theory more 
convenient in analysing the structure of the vacuum.
The calculation of the vacuum charge involves the 
difference between two divergent quantities. The vacuum 
charge may or may not appear fractional depending on 
how the difference is calculated. We propose a natural 
way of resolving this ambiguity. This enables us to 
clearly identify the part of the charge that has been 
left out in earlier analyses. When the external 
potential is switched on the total vacuum charge 
remains conserved. It is not fractionalized. There are 
two other related aspects that we attempt to clarify. 
One is concerned with the usage of the continuum limit 
and the other with the regularization of the current 
operator.  

Before examining the relativistic field with the Dirac
equation it would be useful to review the time-honored 
Friedel sum rule for a nonrelativistic Fermi gas with an 
impurity placed in it \cite{3}. This we do in Sec. II
with emphasis on charge conservation. The impurity 
produces a potential, $V(r)$, that acts on the particles 
in the gas. If we choose $V(r)$ arbitrarily, it appears 
as if the charge induced around the impurity can take 
any fractional value. This is related to the usage of 
the continuum-energy limit. The total charge of the gas 
is conserved. 

In Sec. III we examine a model of the Dirac field in 
one dimension. We see a situation very similar to the
one found in Sec. II. In Sec. IV we examine the vacuum 
confined in a one-dimensional ``bag" of a finite radius. 
Even such a finite bag is beset with an ambiguity that 
is related to the regularization of the current operator. 
A summary and discussions are given in Sec. V. We suggest 
an implication regarding the charge renormalization in 
quantum electrodynamics. We use units such that $c=1$ 
and $\hbar=1$ throughout this paper.

\section{Friedel sum rule: Nonrelativistic Fermi gas
with an impurity}
Consider a nonrelativistic Fermi gas confined in a 
sphere of a large radius $R$ (in three dimensions). 
We eventually let $R\to\infty$. When an impurity is 
placed in the gas, it induces a change in the charge
density distribution. The free gas (without the impurity) 
is described by the Schr\"{o}dinger equation
\begin{equation}
(H_0 -E) \phi(E,{\bf r}) =0, \;\;\; 
H_0 ={\bf p}^2 /(2m),
\label{1}
\end{equation}
In the presence of the impurity at the origin we have
\begin{equation}
(H -E') \psi(E',{\bf r}) =0, \;\;\; H=H_0 +V(r),
\label{2}
\end{equation}
where $V(r)$ is the perturbation potential due to the 
impurity. It is a central, localized potential. The 
$\phi(E,{\bf r})$ and $\psi(E',{\bf r})$ both vanish 
at the boundary of the sphere, i.e., $r=R$, and are 
normalized within the sphere. 

The energy levels are discrete. The $E$ and $E'$ are 
different. If it is attractive (repulsive), $V(r)$ 
shifts the energy levels downward (upward). There is 
a one-to-one correspondence between the perturbed and 
unperturbed energy levels. Imagine that we introduce 
the perturbation as $\eta V(r)$ and let $\eta$ increase 
gradually from 0 to 1. Then the $n$-th $E$-level 
adiabatically goes to the $n$-th $E'$-level. We take 
this one-to-one correspondence for granted.

The density of the gas, defined as the deviation from 
that of the free gas, is given by
\begin{equation}
\rho(r) = \sum_{E'<E'_F}|\psi(E',{\bf r})|^2
-\sum_{E<E_F}|\phi(E,{\bf r})|^2,
\label{3}
\end{equation}
where $E_F$ ($E'_F$) is the free (shifted) Fermi energy. 
The total charge, which we define as the deviation from 
that of the free gas, is
\begin{equation}
Q=4\pi\int_0 ^R r^2\rho (r) dr=\sum_{E'<E'_F}-\sum_{E<E_F}.
\label{4}
\end{equation}
When the perturbation is switched on, the total number 
of the particles in the gas remains the same and hence 
$Q=0$. (See, however, the remarks given towards the end
of this section.) We divide each of the density and the 
charge into two parts,
\begin{equation}
\rho (r) = \rho_c (r) + \rho_d (r), 
\;\;\; Q = Q_c + Q_d ,
\label{5}
\end{equation}
\begin{equation}
\rho_c (r)
= \sum_{E'<E_F}|\psi(E',{\bf r})|^2
-\sum_{E<E_F}|\phi(E,{\bf r})|^2,
\label{6}
\end{equation}
\begin{equation}
\rho_d (r)= {\rm sign}(E'_F -E_F)\sum_{\{E_F ,E'_F\}}
|\psi(E',{\bf r})|^2,
\label{7}
\end{equation}
\begin{equation}
Q_c = 4\pi\int_0 ^R r^2\rho_c (r) dr
= \sum_{E'<E_F}-\sum_{E<E_F},
\label{8}
\end{equation}
\begin{equation}
Q_d = 4\pi\int_0 ^R r^2\rho_d (r) dr
={\rm sign}(E'_F -E_F)\sum_{\{E_F ,E'_F \}}.
\label{9}
\end{equation}
The radius $R$ is finite, although very large. The 
energy levels are discrete. The $\sum_E $, which is 
the number of energy levels within the specified range, 
is well-defined and is an integer. Hence the $Q$'s 
defined above are all integers. 

Let the radial part of the normalized partial-wave 
function in the presence of the impurity be 
$\psi_l (E',r)$. In the asymptotic region where $V(r)$ 
is negligible, we obtain
\begin{equation}
\psi_l (E',r) = \left(\frac{1}{2\pi R}\right)^{1/2}
\frac{1}{r}\sin \left[k'r +\eta_l (E') 
-\frac{l\pi}{2}\right],
\label{10}
\end{equation}
where ${k'}^2 /(2m)=E'$. In Eq. (\ref{6}), the 
difference between $|\psi(E',{\bf r})|^2$ and 
$|\phi(E,{\bf r})|^2$ is small at large distances.
Hence the distribution $\rho_c (r)$ is mostly 
concentrated around the impurity. By using the 
Schr\"{o}dinger equation and Eq. (\ref{10}) and by 
taking the continuum limit, that is, by replacing 
$\sum_E$ with $2(2R/\pi)\sum_l (2l+1)\int dk$, one 
can derive the well-known formula \cite{3},
\begin{equation}
 Q_c =\frac{2}{\pi}\sum_l (2l+1)\eta_l (E_F).
\label{11}
\end{equation}
In deriving Eq. (\ref{11}) one does not have to know the 
wave function in the vicinity of the impurity. If we 
choose $V(r)$ arbitrarily, the above $ Q_c$ can take any 
fractional value. This contradicts what we stated at the
end of the preceding paragraph. We will discuss this 
contradiction shortly. For a real electron gas in a metal, 
only if $V(r)$ is chosen judiciously, Friedel's 
self-consistency condition or the Friedel sum rule
\begin{equation}
Q_c =Z
\label{12}
\end{equation}
is satisfied. Here $Z$ is the (excess) charge of the 
impurity and is an integer.

Next let us turn to $\rho_d (r)$. The Fermi momenta
$k_F$ of the free gas and $k'_F$ of the perturbed gas 
are related by $k'_F R + \eta(E'_F)=k_F R$ and
\begin{equation}
 k_F=k'_F -k_F =-\eta_l (E'_F)/R.
\label{13}
\end{equation}
Unlike $k_F$, $k'_F$ depends on $l$. If the impurity 
is introduced adiabatically, the particle population 
in each of the partial waves remains the same. The 
difference $E_F -E'_F$ is very small (because $R$ is
very large). If we replace $\sum_E$ of Eq. (\ref{7}) with
$2(2R/\pi) \Delta k_F \sum_l (2l+1)\int\delta (k-k_F)dk$,
we obtain,
\begin{equation}
\rho_d (r)=-\frac{2}{\pi}\sum_l (2l+1)
\eta_l (E_F){\psi_l}^2 (E_F,r),
\label{14}
\end{equation}
\begin{equation}
 Q_d = -\frac{2}{\pi}\sum_l (2l+1)\eta_l (E_F).
\label{15}
\end{equation}
This confirms $Q = Q_c + Q_d =0 $. To conserve the total 
charge of the gas, it is crucial to include $\rho_d (r)$ 
which stems from the change in the Fermi energy. In the 
asymptotic region where $V(r)$ is negligible, we obtain
\begin{eqnarray}
4\pi r^2 \rho_d (r) &=&-\frac{4}{\pi R}\sum_l (2l+1)
\eta_l (E_F) \nonumber \\
&& \times \sin^2 \left[k_F r +\eta_l (E_F) 
-\frac{l\pi}{2}\right].
\label{16}
\end{eqnarray}
This distribution is diffuse; it spreads more or less 
uniformly over the entire asymptotic region. As 
$R\to\infty$ the density $\rho_{\rm d}(r)$ vanishes but 
$Q_d $ does not. 

We noted below Eq. (\ref{9}) that $Q_c $ and $Q_d $ are 
integers. This is in contradiction with Eqs. (\ref{11}) 
and (\ref{15}) which show that $ Q_c $ and $Q_d$ can be 
nonintegers and their dependence on $V(r)$ is continuous 
and smooth. The origin of this contradiction can be 
traced as follows. In deriving Eqs. (\ref{11}) and 
(\ref{14}), $\sum_E $ is replaced with $(R/\pi)\int dk$. 
This effectively smears out the discrete energies. The 
number of such smeared-out energies in a specified 
energy range becomes fuzzy. In Eq. (\ref{9}) we count 
the number of energy levels in interval $\Delta E_F 
=E'_F -E_F$, which is of the same order of magnitude 
as the spacing of the discrete energy levels. In terms 
of $k$, the level spacing is $\Delta k=\pi/R$ which is 
comparable with $\Delta k_F =-\eta/R$ of Eq. (\ref{13}). 
In such a case, smoothing the energy levels results in a
considerable uncertainty. For the total charge $Q$, we 
count the number of states in interval $\{0, E_F\}$ 
which is much wider than the energy level spacing. This 
is why the smoothing does not affect $Q$ $(=Q_c +Q_d)$
but it obscures each of $Q_c$ and $Q_d$. The fractional 
values that appear through Eqs. (\ref{11}) and (\ref{15}) 
are artificial.

The model that we have examined above is just a 
mathematical model. Let us examine its relevance to a 
real electron gas in a metal. For example, the impurity 
can be a Zn ion of charge +2 which replaces a Cu ion 
of charge +1 in a metal lattice. The excess charge is 
$Z=1$. In a real electron gas we have to remember the 
electron-electron interaction which is repulsive. This 
pushes the electrons in the diffuse distribution 
towards the surface of the metal. No diffuse 
distribution remains inside the metal. (This is a 
well-known result of electrostatics.)

We have assumed that the impurity potential $V(r)$ 
is switched on adiabatically. This, however, is not 
what actually takes place in the real electron gas. 
As pointed out by Friedel, when one of the ions is 
replaced by an impurity of excess charge $Z$, additional 
electron(s) are added to the system at the same time. 
The charge of the added electrons exactly cancels $Z$ 
and the entire system, including its background ions, 
remains neutral \cite{3,4}. The Fermi energy of the gas 
with the added electron(s) becomes the same as the free 
Fermi energy $E_F$. If the electron-electron and 
electron-ion interactions are taken into account in a 
self-consistent manner, e.g., by means of the 
Hartree-Fock method, $V(r)$ will adjust itself such 
that $ Q_c =Z$ is satisfied and the impurity charge 
is screened. Since $Z$ is an integer, no fractional 
charge appears. 

We raised a question regarding the continuum limit of
the energy levels. We, however, do not propose to 
dismiss the $Q_c$ and $Q_d$ obtained in the continuum 
limit. It is conceivable that Eqs. (\ref{11}) and 
(\ref{15}) are practically acceptable or even preferable. 
The reason is as follows. Imagine that $V(r)$ is varied 
continuously and one of the $E'$ levels crosses $E_F$. 
Then $Q_c$ of Eq. (\ref{6}) suddenly changes by 
$2(2l+1)$ where $l$ is the angular momentum of the $E'$ 
level. This is probably unrealistic. The energy levels
near the Fermi energy may be only partially occupied.
The fuzziness that is introduced by smearing the energies
may be appropriate in such a situation. It is well-known 
that the Friedel sum rule (\ref{12}) is useful \cite{3,4}.

\section{Dirac field in one dimension}
We examine the vacuum charge of a Dirac field in one 
dimension. We start with the Dirac equation for a 
particle in a given external potential in single particle 
quantum mechanics. We compare the free  and perturbed 
systems. The Dirac equation for the free system is
\begin{equation}
(H_0 -E)\phi(E,x) =0, \;\;\; H_0 = \alpha p + \beta m,
\label{17}
\end{equation}
where $p=-id/dx$, $\alpha$ and $\beta$ are the $2\times 2$
Dirac matrices and $m$ is the mass of the particle. In 
explicit calculations we use $\alpha=\sigma_2$ and 
$\beta=\sigma_3$. The Dirac equation for the perturbed 
system is
\begin{eqnarray}
&&(H -E')\psi(E',x) =0, \nonumber \\ 
&&H=H_0 + U(x), \;\;\; U(x)=\beta S(x) + V(x),
\label{18}
\end{eqnarray}
where $S(x)$ is a Lorentz scalar and $V(x)$ the zero-th 
component of a Lorentz vector. 

If $V(x)=0$, there is a symmetry between the positive and 
negative energy spectra: $\psi(E',x)$ and $\psi(-E',x)$ 
are related by $\psi(-E',x)=\alpha\beta\psi(E',x)$. 
Because there cannot be energy-degeneracy in one dimension, 
this symmetry may seem to exclude the possibility of a zero 
mode. A zero mode, however, is allowed provided that $S(x)$ 
has a topology such that $m+S(\infty)$ and $m+S(-\infty)$ 
are of opposite signs \cite{1,4a}. Such a situation arises 
with a kink soliton background. The zero mode remains no 
matter how $S(x)$ is modified as long as this topology 
is maintained. If $V(x)\neq 0$, the symmetry between the 
positive and negative energy spectra is broken. But the 
zero mode can still exist as long as 
$[m+S(\infty)][m+S(-\infty)]< 0$ and $|V(x)|<|S(x)|$ 
when $x \to \pm \infty$. The zero mode is stable against
the variation of $V(r)$. This is the type of zero mode
that underlies half-integral charges \cite{1}.

In this paper we assume that $S(x)$ has no such topology 
as described above. To be more explicit, we assume that 
$[m+S(\infty)][m+S(-\infty)]\geq 0$.
Under this assumption for $S(x)$, it is still possible 
to have a zero mode. This can be done by carefully 
adjusting $V(x)$ such that one of the eigenvalues 
becomes zero. This zero mode is obviously unstable 
against the variation of $V(x)$. Its energy eigenvalue 
can deviate from zero even if the variation of $V(x)$ 
is very slight. We may say that the existence of this 
zero mode is accidental. This is in contrast to the 
zero mode that we described in the preceding paragraph. 
In any case we assume that our Dirac Hamiltonian $H$ 
has no zero mode.

The $H$ of Eq. (\ref{18}) can be transformed into 
different forms. An example is,
\begin{eqnarray}
H_f &\equiv& e^{i\alpha f(x)}He^{-i\alpha f(x)} \nonumber \\
&=&\alpha p + V(x) -\frac{df(x)}{dx} \nonumber \\
&+&\beta [m+S(x)][\cos 2f(x) -i\alpha \sin 2f(x)].
\label{19}
\end{eqnarray}
The term with $-i\beta\alpha=-i\sigma_3 \sigma_2 =-\sigma_1$ 
is a pseudoscalar potential. Potential $V(x)$ can be 
eliminated by choosing $f(x)$ such that $df(x)/dx=V(x)$. 
If $m+S(x)=0$, then the transformed Hamiltonian $H_f$ simply 
becomes $H_0=\alpha p$, which is of course solvable. We
will make use of this transformation in the next section.
In this section we assume that the mass is nonzero. 

Suppose we eliminate $V(x)$ in $H_f$ by means of 
$df(x)/dx=V(x)$. It is understood that $m+S(\infty)$ and 
$m+S(-\infty)$ are of the same sign. Assume that 
$f(\infty)f(-\infty)<0 $. Then the peudoscalar part of 
the potential of $H_f$, which is proportional to 
$i\alpha\beta\sin 2f(x)$, has a topology, that is, its 
asymptotic values for $x \to \pm \infty$ are of opposite 
signs. If we go back to $H$, however, there is no such 
topology because $df(x)/dx$ has the same sign at 
$x \to \pm \infty$. Hamiltonians of the form of $H_f$ are 
extensively used in the literature \cite{2} and one might 
have the impression that the fractional charge associated 
with such $H_f$ is due to its topology. This topology, 
however, is a trivial one in the sense that it can be 
transformed away. Let us emphasize that the topology 
carried by the Lorentz scalar potential remains the
same in transformation (\ref{19}). The topology of
$S(x)$ that we mentioned in the paragraph below 
Eq. (\ref{18}) cannot be transformed away. 

We return to Hamiltonian $H$ of Eq. (\ref{18}). We assume 
that the system is confined within a box of large radius 
$R$, i.e., $r=|x|<R$. We do this by assuming an infinite 
square-well potential of the Lorentz scalar type, which 
is $\infty$ for $r>R$. It is understood that this 
square-well potential is a part of $S(x)$. The wave 
function has to satisfy a boundary condition at $r=R$; 
see Eq. (\ref{42}) of the next section. Since $R$ is 
finite, the energy levels are discrete. We let 
$R\to\infty$ in the end. (We keep $R$ finite in Sec. IV.)
The wave function vanishes for $r>R$. Therefore, $S(x)$ 
and $V(x)$ for $r>R$ do not appear in the following 
calculations. We assume that $S(x)$ and $V(x)$ within 
the box are both localized around the origin. In this 
way we can see a parallelism between this and the 
preceding sections more easily. In the following 
calculations, however, the $S(x)$ and $V(x)$ within 
the box can be chosen arbitrarily. 

We can find a one-to-one correspondence between the free 
energy levels and the perturbed ones as we did in Sec. II. 
For simplicity we assume that positive $E$s correspond 
to positive $E'$s. In other words, the perturbation does 
not change the sign of the energy. The vacuum is such 
that the negative energy levels are all occupied. The 
number of the negative energy levels is of course infinite. 
We, however, first assume that the negative energy 
levels are filled only down to a certain Fermi energy
$E_F <0$ for the free vacuum and similarly to $E'_F <0$ 
for the perturbed vacuum. Eventually we let the Fermi 
energies tend to $-\infty$. Unlike the Fermi energy of 
the nonrelativistic Fermi gas, the Fermi energy of the 
vacuum is a mathematical device. We need such a device 
so that we can clearly keep track of the number of 
particles in the vacuum.   

The density and the charge of the perturbed vacuum are
respectively defined as the deviations from those of 
the free vacuum,
\begin{equation}
\rho(x) = \sum_{0>E'>E'_F}|\psi(E',x)|^2
-\sum_{0>E>E_F}|\phi(E,x)|^2, 
\label{20}
\end{equation}
\begin{equation}
Q =\int_{-R} ^R \rho (x) dx 
=\sum_{0>E'>E'_F} -\sum_{0>E>E_F}.
\label{21}
\end{equation}
The one-to-one correspondence between the free and 
perturbed energy levels implies $ Q=0$. When the 
perturbation is switched on, the total charge 
remains conserved. 

Exactly in the same way as we did in Sec. II we divide
each of the density and the charge into two parts,
\begin{equation}
\rho (x) = \rho_c (x)+ \rho_d (x), \;\;\;
 Q = Q_c + Q_d ,
\label{22}
\end{equation}
\begin{equation}
\rho_c (r) = \sum_{0>E'>E_F}|\psi(E',x)|^2 
-\sum_{0>E>E_F}|\phi (E,x)|^2,
\label{23}
\end{equation}
\begin{equation}
\rho_d (r)
= {\rm sign}(E_F -E'_F )\sum_{\{E_F ,E'_F \}}|\psi(E',x)|^2,
\label{24}
\end{equation}
\begin{equation}
Q_c = \int_{-R} ^R \rho_c (x)dx
= \sum_{0>E'>E_F}-\sum_{0>E>E_F},
\label{25}
\end{equation}
\begin{equation}
Q_d = \int_{-R} ^R \rho_d (x)dx
= {\rm sign}(E_F -E'_F )\sum_{\{E_F ,E'_F \}}.
\label{26}
\end{equation}
The $Q$ can be rewritten as 
\begin{equation}
Q = \frac{1}{2}\left(\sum_{0>E'>E'_F}
-\sum_{0<E'<E''_F}\right).
\label{27}
\end{equation}
Imagine the positive-energy level of $E=|E_F|$ of the 
free system. The $E''_F >0$ of Eq. (\ref{27}) is the 
energy level (of the perturbed system) that corresponds 
to $|E_F|$. Figure 1 schematically shows how the Fermi 
energy shifts when the perturbation is attractive. In 
general $E''_F \neq -E'_F$. In deriving Eq. (\ref{27}) 
we used
\begin{equation}
\sum_{0>E'>E'_F}+\sum_{0<E'<E''_F}=2\sum_{0>E>E_F},
\label{28}
\end{equation}
which is based on the one-to-one correspondence between
$E$s and $E'$s. 
Furthermore the $ Q_c$ can be written as
\begin{equation}  
 Q_c = \frac{1}{2}\left(\sum_{0>E'>E_F}
-\sum_{0<E'<|E_F |}\right).
\label{29}
\end{equation}
The $ Q_c$ is a measure of the asymmetry between the 
positive and negative energy spectra. 

\begin{center}
\unitlength 0.8mm
\linethickness{0.4pt}
\begin{picture}(85,60)
\put(28,10){\line(1,0){28}}
\put(25,28){\line(1,0){33}}
\put(28,46){\line(1,0){28}}

\put(28,05){\line(1,0){3}}
\put(33,05){\line(1,0){3}}
\put(38,05){\line(1,0){3}}
\put(43,05){\line(1,0){3}}
\put(48,05){\line(1,0){3}}
\put(53,05){\line(1,0){3}}

\put(28,41){\line(1,0){3}}
\put(33,41){\line(1,0){3}}
\put(38,41){\line(1,0){3}}
\put(43,41){\line(1,0){3}}
\put(48,41){\line(1,0){3}}
\put(53,41){\line(1,0){3}}

\put(42,9){\vector(0,-1){3}}
\put(42,45){\vector(0,-1){3}}

{\small
\put(16,10){\makebox(0,0)[cc]{$E_F <0$}}
\put(16,28){\makebox(0,0)[cc]{$E=0$}}
\put(16,46){\makebox(0,0)[cc]{$|E_F |$}}
\put(68,05){\makebox(0,0)[cc]{${E'}_F <0$}}
\put(68,41){\makebox(0,0)[cc]{$E''_F >0$}}
}
\end{picture}

{\small 
{\bf Fig. 1} \ \ 
Unperturbed and perturbed Fermi energies.}
\end{center}
\vskip 0.5cm


When $E_F$, $E'_F$ and $R$ are finite, the $\sum_E$'s 
are all well-defined. The above $Q$'s are all integers. 
When $E_F \to -\infty$, $E'_F \to -\infty$ and 
$R \to \infty$, the $Q$'s remain as the same integers. 
In Sec. II and so far in this section we have 
distinguished the unperturbed and perturbed energies, 
using different notations $E$ and $E'$. In the 
following, when the distinction between $E$ and $E'$ is 
unimportant or is clear from the context, we may denote 
both of them as $E$ for notational brevity. 

The $\rho_c (x)$ is the one that has been extensively 
examined in the literature, for example, in Refs. 
\cite{5,6}. According to those analyses it is given by
\begin{equation}
 Q_c = -\frac{1}{\pi}\eta(-\infty),
\label{30}
\end{equation}
where we have taken the limit of $E_F \to -\infty$. The 
phase shift $\eta (E)$ pertains to the transmission 
coefficient $T(E)$. In the Born approximation the phase 
shifts are given by 
\begin{equation}
\eta (E) = -\frac{1}{k}\int^R _{-R}[EV(x)+mS(x)]dx,
\label{31}
\end{equation}
which becomes exact when $|E| \to \infty$.
Thus we arrive at 
\begin{equation}
 Q_c =-\frac{1}{\pi} \int^R _{-R} V(x)dx.
\label{32}
\end{equation}
The $S(x)$ has no effect on $ Q_c$. It follows from Eq. 
(\ref{31}) that $\eta(\infty)=-\eta(-\infty)$ and 
Eq. (\ref{30}) can be written as 
$ Q_c = (1/2\pi)[\eta(\infty)-\eta(-\infty)]$.
This shows that $ Q_c$ is a measure of the asymmetry 
between the positive and negative energy phase shifts. 
Let us add that, in the nonrelativistic version 
of the present model, the phase shifts in the Born 
approximation are $\eta (E) = -(m/k)\int^R _{-R}V(x)dx$. 
Unlike its relativistic counterpart, the nonrelativistic 
$\eta (E)$ vanishes as $E\to \infty$.

No attention seems to have been paid to $\rho_d (r)$ in 
the literature so far. In order to see similarity to 
the calculation of Sec. II clearly, let us assume that 
$U(x)$ of Eq. (\ref{18}) is an even function of $x$ so 
that parity is a good quantum number. There are two 
partial waves, one with even parity and the other with 
odd parity \cite{7}. The transmission coefficient $T(E)$ 
is related to the partial wave phase shifts $\eta_\pm$ 
by $T=[e^{2i\eta_+}+e^{2i\eta_-}]/2$ and hence
\begin{equation}
\eta (E) = \eta_+ (E) +\eta_- (E).
\label{33}
\end{equation}
In the asymptotic region where $U(r)$ is negligible, 
the perturbed wave function of energy $|E|>m$ and 
positive parity takes the following form,
\begin{equation}
\psi_+ (E,x) \nonumber \\
=\sqrt{\frac{E+m}{2RE}} \left( \begin{array}{c}
\cos(kr+\eta_+ ) \\
\frac{k}{E+m}\hat{x}\sin(kr+\eta_+ )
\end{array} \right),
\label{34}
\end{equation}
where $k=\sqrt{E^2 -m^2}$ and $\hat{x}=x/r$. Suffix 
$+$ of the phase shift $\eta_+ (E)$ refers to positive 
parity. There may be bound states with $|E|<m$. For 
negative parity we obtain
\begin{equation}
\psi_- (E,x)
=\sqrt{\frac{E-m}{2RE}} \left( \begin{array}{c}
\frac{k}{E-m}\hat{x}\sin(kr+\eta_- ) \\
\cos(kr+\eta_- ) 
\end{array} \right).
\label{35}
\end{equation}
The $\phi_\pm (E,x)$ are obtained by dropping $\eta_\pm $. 

The Fermi momenta before and after perturbation $U(x)$ is 
switched on, $k_F$ and $k'_F$, are related again by 
Eq. (\ref{13}). By taking the continuum limit, we obtain
\begin{equation}
\rho_d (x) =\frac{1}{\pi}\sum_{l=\pm}
\eta_l (E_F )|\psi_l (E_F ,x)|^2 .
\label{36}
\end{equation}
By integrating this with respect to $x$ and letting 
$E_F \to -\infty$ we obtain
\begin{equation}
 Q_d =\frac{1}{\pi}\eta(-\infty)
=\frac{1}{\pi} \int^R _{-R} V(x)dx.
\label{37}
\end{equation}
In the asymptotic region where $U(x)$ is negligible,
in the limit of $E_F \to -\infty$, we find that
$|\psi_+ (E,x)|^2 \to 1/(2R)$ and hence 
$\rho_d (x) \to (1/\pi R)\eta (-\infty)$.

The remark given in Sec. II regarding $Q_c$ and $Q_d$ 
also applies to the $Q_c $ and $Q_d $ obtained above. 
As a model, $V(x)$ can be chosen arbitrarily. Then 
$Q_c$ of Eq. (\ref{32}) and $Q_d$ of Eq. (\ref{37}) can
take any fractional values. As we noted below Eq. 
(\ref{29}), however, the $Q$'s of Eqs. (\ref{25}) and
(\ref{26}) are integers. The fractional values of 
$ Q_c$ of Eq. (\ref{32}) and $Q_d$ of Eq. (\ref{37}) 
are mathematical artifacts that stem from replacing 
discrete summation $\sum_E$ with the integration 
$(R/\pi)\int dk$ over the smoothed energy levels. 

At the end of Sec. II we suggested a practical 
justification for $Q_c$ of Eq. (\ref{11}) of the 
nonrelativistic electron gas in a metal. Such 
justification is untenable for the present 
relativistic model in which the Fermi energy is 
a purely mathematical device. Towards the end of 
this section we argue that, although they should 
not be taken literally, these artificial results do 
no harm in real physics.

Let us briefly discuss an explicit example. Assume that
\begin{equation}
S(x)=0, \;\;\; V(x) = -2\lambda \delta(x).
\label{38}
\end{equation}
This is a special case $(\lambda_1 =\lambda_2 =\lambda)$ 
of the model that was examined in Ref. \cite{8}. This 
model is related to the model examined by MacKenzie 
and Wilczek \cite{5} by transformation (\ref{19}).
In dealing with the $\delta$-function potential, we 
start with a square-well potential and after solving 
the Dirac equation we take the narrow-width limit of 
the square well-potential. In this connection see Ref. 
\cite{9}. The phase shifts are given by 
\begin{equation}
\tan\eta_\pm (E) = \frac{E\pm m}{k}\tan \lambda.
\label{39}
\end{equation}
If $g<0$ there is a bound state of positive parity. 
Its energy is
\begin{equation}
E = m\cos(2\lambda).
\label{40}
\end{equation}
The Dirac matrices used in Ref. \cite{8} are 
$\alpha=\sigma_1$ and $\beta=\sigma_3$, which can be 
transformed to $\alpha=\sigma_2$ and $\beta=\sigma_3$ 
that we are using by the unitary transformation
$\exp(i\pi\sigma_3 /4)$. The wave function of the 
bound state is
\begin{equation}
\psi(x)=\sqrt{\frac{\kappa (E+m)}{2m}}
\left( \begin{array}{c} 
1 \\ \frac{\kappa}{E+m}\hat{x} \end{array}
\right) e^{-\kappa r},
\label{41}
\end{equation}
where $\kappa = \sqrt{m^2 -E^2}$. If $g>0$ there is a 
bound state of negative parity but we do not delve 
into such details in this paper. The $\rho_c (x)$ was
worked out explicitly in Ref. \cite{8}. The summation 
$\sum_E $ was done as $(R/\pi)\int dk$. The $\rho_c (x)$ 
is indeed concentrated around the origin. It decays 
like $e^{-2mr}$. If we let $m \to 0$, $\rho_c (x)$ 
becomes uniform. By carrying out the integration 
$\int \rho_c (x)dx$ explicitly, one obtains $ Q_c$ of 
Eq. (\ref{32}). The $\rho_d (x)$ was not considered in 
Ref. \cite{8} but it is given by Eq. (\ref{36}). It is 
uniform. Let us add that a model in which the 
$\delta$-function potential is replaced by a square-well 
potential has been also examined \cite{10}. The square 
well-version is related to the model of Ref. \cite{11} 
by transformation (\ref{19}). This ends the discussion
of the example.

As we pointed out already the fractional charges of 
Eqs. (\ref{32}) and (\ref{37}) are mathematical 
artifacts. Still let us speculate on what happens if we 
accept the fractional charges literally. In the model 
that we have examined, the particles individually 
interact with the external potential but there is no 
interaction between the particles. We assumed that the 
external potential can be chosen arbitrarily. In real 
physical systems, however, the particles generally 
interact with each other. When the inter-particle 
interaction is (approximately) eliminated in a 
self-consistent manner, the potential that emerges is 
not an arbitrary one. If the particle of the field is 
the electron in a metal, the remarks given at the end 
of Sec. II apply. The repulsion between electrons 
pushes the electrons in the diffuse distribution 
towards the edge at $r=R$. Such a charge accumulated 
at the edge can easily move on to another object that 
comes into contact with the metal. For $ Q_c$, the 
Friedel sum rule (\ref{12}) will be satisfied. Thus 
the self-consistency prevents fractionization of the 
vacuum charge.

Our model can be regarded as a one-dimensional 
simulation of the relativistic nuclear shell model. 
In that case we have a number of nucleons in the 
positive energy levels, but let us focus on the vacuum 
effect. According to the Dirac phenomenology of nuclear 
physics, the relativistic potential of the nuclear 
shell model is of the form of $U(r)=\beta S(r) +V(r)$ 
\cite{12}. For medium to heavy nuclei, $S(r)$ and $V(r)$ 
are well represented by the Wood-Saxon form. Inside the 
nucleus $S+V \approx -50$ MeV and $-S+V\approx 800$ MeV. 
The potential acting on the lower component of the 
Dirac wave function is $-S+V$, which is almost as large 
as the nucleon rest mass of 940 MeV. Consequently 
relativistic effects can be appreciable even at very 
low energies. If we assume $V=400$ MeV, we obtain 
$ Q_c = 400(2 R_0 /\pi)$ MeV$\cdot$fm where 
$R_0 \approx (1.2)A^{1/3}$ is the radius in fm of the 
nucleus of mass number $A$. If we assume $A=100$, 
for example, we find $R_0 \approx 5.6$ fm and 
$ Q_c =-Q_d \approx -7.2$. Approximately 7 nucleons
are moved from the concentrated region to the diffuse 
distribution. This number is not negligibly small.

In our model calculation the diffuse distribution is 
uniform in the asymptotic region (outside the range of 
the shell model potential). Unlike the electron case,
the interaction between the nucleons is attractive. When 
the nucleons are treated in a self-consistent manner, 
like the Hartree-Fock method, the interaction between 
the nucleons will produce an attractive potential and the 
diffuse distribution will be pulled back to a finite 
region. It may become something like a nuclear halo. 
The nucleons, including those in the diffuse 
distribution, will form a nucleus of a finite size. 
The total charge of the nucleus is an integer. It is 
an interesting challenge to do such a self-consistent 
calculation including the vacuum effect. If the nucleons 
in the diffuse distribution somehow escaped to infinity, 
the nucleus would be left with $ Q_c$. If this $ Q_c$
takes a fractional value, the nuclear charge becomes
fractional. This would lead to bizarre consequences. 
We know that this is not the case in reality. 

Finally let us mention super-heavy quasimolecules that
could be formed in a collision of two very heavy ions. 
This subject has been extensively discussed by Greiner 
et al. \cite{13}. When the strength of the Coulomb 
potential due to the merging nuclei exceeds a certain 
critical value, the lowest positive energy level of the
electron dives into the negative energy sea. This may 
lead to the creation of an electron-positron pair. 
Suppose that the vacuum charge is fractionalized in 
this process. Then can a fraction of the pair be 
created? This does not seem to make sense.

\section{Vacuum confined in a finite bag}
We examine a one-dimensional version of the so-called 
MIT bag model \cite{14}. We also discuss the chiral bag 
model \cite{15} towards the end of this section. We 
keep the radius $R$ of the bag finite. In Sec. III we 
divided the density into two parts, concentrated and 
diffuse. Such division is not very meaningful when $R$ 
is not very large. In Secs. II and III, we pointed out 
that replacing $\sum_E $ with $(R/\pi)\int dk$ leads 
to fractional charges. When $R$ is finite and not very 
large, one may think that there is no room for such 
complication and no problem arises regarding charge 
conservation because the charge cannot escape to 
infinity. Nevertheless, the vacuum charge may be 
fractionalized in the bag, depending on how it is 
calculated. 

The Hamiltonian of the bag model is the same as that
of Sec. III. We assume an infinite square well 
potential for $r>R$ so that the particle is confined 
to $r<R$. This leads to the boundary condition at $r=R$, 
\begin{equation}
(1-i\alpha\beta \hat{x})\psi(x)
=(1+\sigma_1 \hat{x})\psi(x) =0,
\label{42}
\end{equation}
which is equivalent to Eq. (14) of Ref. \cite{16}.
This condition makes the scalar density 
$\psi^\dagger \beta \psi $ vanish at $r=R$. 

We begin with the simple case of $m=0$. The solutions
of the free Dirac equation can be classified in terms 
of parity. For the unperturbed system with $H_0$, we 
obtain
\begin{equation}
\phi_+ (x) =N\left( \begin{array}{c}
\cos kx \\ -\sin kx \end{array}\right),
\label{43}
\end{equation}
\begin{equation}
\phi_- (x) =N\left( \begin{array}{c}
\sin kx \\ \cos kx \end{array}\right) 
= -i\alpha\phi_+ (x),
\label{44}
\end{equation}
where suffix $\pm$ refers to parity and $N=1/\sqrt{2R}$ 
is the normalization factor. The density $|\phi (E,x)|^2
= 1/(2R)$ is a constant. The $k$ and $E=\pm k$ are 
determined by the boundary condition (\ref{42}). The 
positive energy eigenvalues are given by
\begin{equation}
{E_n}^{(\pm)} =\pm \frac{(2n-1)\pi}{4R}, 
\;\;\; n=1,2,3, \dots\, ,
\label{45}
\end{equation}
where superscript $(\pm)$ refers to the sign of the
energy. All energy levels are equally spaced with 
interval $\pi/(2R)$.

We now examine the perturbed system with the 
$H=H_0 +V(x)$. We assume $S(x)=0$ for simplicity.
The $H$ and $H_0$ can be related by transformation 
(\ref{19}). The following two wave functions satisfy 
the perturbed Dirac equation,
\begin{equation}
\psi_+ (x) =N\left( \begin{array}{c}
\cos [k'x -f(x)] \\ -\sin [k'x -f(x)] \end{array}\right),
\;\;\; \frac{df(x)}{dx} = V(x),
\label{46}
\end{equation}
and $\psi_- (x)=-i\alpha \psi_+ (x)$. In order for the
above to satisfy the Dirac equation, $V(x)$ does not 
have to be an even function of $x$. If $V(x)$ is not an
even function, parity is not a good quantum number. We
still retain the suffixes $\pm$ because the above 
solutions are respectively related to the free solutions
$\phi_\pm (x)$. The above is a generalization of the
solution of Ref. \cite{17} in which $V(x)=\lambda x$ 
was assumed.
 
The perturbed density is $|\psi (E',x)|^2 =1/(2R)$. 
Imposing the condition (\ref{42}) we find that the 
perturbed energies and the energy shift,
\begin{equation}
\Delta E =E'-E=\frac{1}{2R}\int^R _{-R} V(x)dx. 
\label{47}
\end{equation}
This $\Delta E$ has two remarkable aspects. 
(i) The $\Delta E$ is of the form of the first-order 
perturbation; yet it is exact. This is so no matter 
how strong $V(x)$ is. Higher order terms are absent. 
If $V(x)$ is an odd function of $x$, then $\Delta E=0$.
(ii) This energy shift $\Delta E $ applies to all 
energy levels, that is, all energy levels are shifted 
by exactly the same amount. 

In the vacuum all negative energy levels are occupied.
Start with the free vacuum and switch on perturbation
$V(x)$. Assume that $V(x)$ is attractive so that 
$\Delta E<0$. Recall that all energy levels are equally 
spaced with separation $\pi/(2R)$. If $|\Delta E| 
<\pi/(2R)$, the lowest positive energy remains positive 
and hence the number of the negative energy levels 
remains the same. There is no change in the vacuum 
charge, i.e., $ Q=0$. If $\pi/(2R)<|\Delta E| <\pi/R$, 
the lowest positive $E$ becomes negative $E'$. Therefore
the number of the negative energy levels increases by
one. This new negative energy level is empty. If
we fill this level by hand, the charge of the new 
vacuum becomes one unit greater than that of 
the free vacuum. No fractional charge appears.

Suppose that $|\Delta E|\gg\pi/(2R)$. Then a large 
number of positive energy levels move into negative 
energy sea. The number of such levels can be estimated 
as
\begin{equation}
-\frac{\Delta E}{\pi/(2R)} 
= -\frac{1}{\pi}\int^R _{-R} V(x)dx.
\label{48}
\end{equation} 
We have put the negative sign in the left hand side 
of Eq. (\ref{48}) because the number should be positive 
when $\Delta E $ is negative. It is interesting that 
the above result is consistent with the $ Q_{\rm c}$ of
Eq. (\ref{32}). However, the number of the levels 
that goes from positive to negative should always be an 
integer. Equation (\ref{48}) is only a fuzzy estimate 
of the discrete number.

Equation (\ref{27}) for $ Q$ also holds in the present 
case. When $E'_F \to -\infty$ and $E''_F \to \infty$, 
$ Q$ is the difference of two divergent series. The 
difference depends on how the summation is done. We 
propose to do this in the same way as we did in Sec. III. 
We first keep $E'_F$ and $E''_F $ finite. Then the 
one-to-one correspondence between the $E$s and $E'$s 
dictates that $ Q=0$. This remains so in the limit of 
$E'_F \to -\infty$ and $E''_F \to \infty$. This we 
believe is a most natural prescription for dealing 
with the series.

There is another prescription that has been widely 
used \cite{18}. This is to rewrite $ Q$ as
\begin{equation}
 Q = \frac{1}{2}\lim_{s\to +0}
\left(\sum_{0>E'}e^{-sE'}-\sum_{0<E'}e^{sE'}\right),
\label{49}
\end{equation}
where the summation is extended to infinity. (No
Fermi energy is assumed.) This prescription is based 
on the so-called split-point regularization of the 
density operator in quantum field theory \cite{19}. 
Let us apply this method to our bag model. With 
${E'}^{(\pm)} _n={E_n}^{(\pm)} +\Delta E$ where
${E_n}^{(\pm)}$ is that of Eq. (\ref{50}) we obtain
\begin{equation}
\sum_{0>E'}e^{-sE'}=
\frac{e^{s\Delta E}\exp{\left(\frac{\pi s}{4R}\right)}}
{1-\exp{\left(-\frac{\pi s}{2R}\right)}},
\label{50}
\end{equation}
and similarly for the other summation. By taking the
limit of $s\to +0$ and using Eq. (\ref{47}) we obtain
\begin{equation}
 Q = -\frac{1}{\pi}\int^R _{-R}V(x)dx.
\label{51}
\end{equation}

We have three remarks to make on the above result. \\
(i) Equation (\ref{51}) follows no matter how small 
$R$ is and hence no matter how large the level spacing 
is. If the level spacing is very large and $V(x)$ is 
not very strong, all positive energy levels remain 
positive. The number of the negative-energy levels 
does not change. Even then Eq. (\ref{51}) implies that 
the vacuum charge changes, which we find unphysical. 
The current is zero everywhere in stationary states. 
The charge cannot escape from the bag and hence it
must be conserved. \\
(ii) Imagine that the energy levels are continuous and
replace the sum over $n=1,2,3,\dots$ of Eq. (\ref{49}) 
with an integral $\int^\infty dn$. We still obtain
Eq. (\ref{51}). (This result is independent of the 
lower end of the integral.) In this sense the 
prescription (\ref{49}) is equivalent to smearing out
the discrete energy levels. It is interesting that
this smearing leads to the same fractional charge as
that of $ Q_c $ of Eq. (\ref{32}) which is also
related to the smearing of the energy levels as we
discussed in detail in Sec. II. We think this is a 
mathematical artifact. \\
(iii) Divide $ Q$ into $ Q_c $ and $ Q_d $ as we did 
in Sec. III. (This division is not interesting in the
sense that, when $m=0$, $\rho_c$ and $\rho_d$ are both
diffuse.) If we use Eq. (\ref{49}), we find that 
$ Q_d$ disappears. The prescription (\ref{49}) is 
equivalent to ignoring the change in the Fermi energy 
in our proposed scheme. In the case of Sec. III, 
$Q_c$ and $Q_d$ may both be fractionalized but they 
cancel each other, $Q =Q_c +Q_d =0 $. In the present 
case, fractionalized $ Q_c $ remains because $ Q_d $ 
is missing. The total charge is not conserved. This 
is unphysical.
 
So far we have assumed $m=0$. We now assume that
$m>0$. The Dirac equation can be solved exactly for
$H_0$ but not for $H$ in general. As $|E'|$ becomes 
much larger than $m$, however, the effect of the 
finite mass becomes negligible.  For such $E'$s the 
energy shift is given by Eq. (\ref{47}). The one-to-one 
correspondence between the $E$s and $E'$s implies 
$ Q=0$. Here we are again assuming that positive 
$E$'s correspond to positive $E'$s, for simplicity. 
No fractional vacuum charge appears.  

Let us see what happens if we use prescription 
(\ref {49}) when $m>0$. The $ Q$ is essentially 
determined by the energy shift for very large $E'$s. 
This situation is similar to the one we saw in Sec. 
III where $ Q$ was determined by $\eta (\pm\infty)$. 
It is not difficult to see by using Eq. (\ref{49}) 
for the cases of $m>0$ and $m=0$ and remembering 
that the $E'$ becomes independent of $m$ as 
$|E'| \to \infty$, that
\begin{equation}
 Q (m>0) - Q (m=0)=0.
\label{52}
\end{equation}  
Hence Eq. (\ref{51}) also holds even when $m>0$. 

We now turn to the chiral bag model \cite{15,18,20}. 
Fractionization of the vacuum charge has been 
extensively discussed for the chiral bag model, but 
never for the MIT bag model that we are using. This 
may give the wrong impression that fractionization 
is a peculiar feature of the chiral bag model. As 
far as the vacuum charge associated with the bag is 
concerned, however, there is no essential difference 
between the two types of the bag models. 

It is instructive to examine the one-dimensional 
version of the chiral bag model that was proposed by 
Zahed \cite{20}. The quark is assumed to be massless.
The quark wave function $q(x)$ for a stationary state 
obeys the Dirac equation
\begin{equation}
(\alpha p -E)q(x) =0,
\label{53}
\end{equation}
inside the bag and the boundary condition at 
$x=\pm R$
\begin{equation}
[\cos\theta (x)-i\alpha \sin\theta (x) 
+\sigma_3 \hat{x}] q(x) =0,
\label{54}
\end{equation}
where $\alpha =\sigma_2$ and $\theta (x)$ is a function 
that pertains to the soliton with which the quark 
interacts. If we write $q(x)$ as
\begin{equation}
q(x) = \exp \{(i/4)[2\theta(x)-\pi ]\alpha \} \psi (x),
\label{55}
\end{equation}
Eq. (\ref{53}) becomes
\begin{equation}
\left[\alpha p 
+\frac{1}{2}\frac{d\theta(x)}{dx} -E \right]\psi (x) =0, 
\label{56}
\end{equation}
and the boundary condition (\ref{54}) is reduced to
Eq. (\ref{42}). As far as the quarks in the bag are 
concerned, Zahed's model is exactly equivalent to 
the one-dimensional MIT bag model with external 
potential $V(x) = (1/2)d\theta(x)/dx$ and the quark 
mass $m=0$. The energy spectrum and the energy shift 
that Zahed obtained are the same as those of of the 
MIT bag model. This situation also holds in three 
dimensions, that is, the chiral bag model can be 
transformed into the MIT bag model with an external 
potential.

From our point of view, there is no anomalous, 
fractional baryon number associated with the bag. 
We can still have a soliton which interacts with the 
bag, and the bag plus the soliton can be interpreted 
as a baryon. The baryon number carried by the bag, as 
we see it, is an integer. There is no need to invoke 
the interpretation that a part of the baryon number 
is carried by the soliton as is done in the chiral 
bag model \cite{20,21}.

We should add that not all the people who worked with
the chiral bag model assumed that the baryon number
in the bag is fractional. In the models for baryons 
developed in Ref. \cite{22}, the baryon number of the
chiral bag is unity. Examining the vacuum charge of a
model similar to that of Zahed, Banerjee argued that 
``there is nothing wrong about departing from the 
split-point form" in regularizing the current operator 
\cite{23}. We concur with him in this respect. 
Banerjee then suggested a regularization scheme such 
that, even in the presence of spectrum asymmetry that 
is induced by an external perturbation, one uses 
$e^{\pm sE}$ instead of $e^{\pm sE'}$ for $Q$ of 
Eq. (\ref{49}). This ad hoc scheme leads to $Q=0$. 
We believe, however, the scheme with the Fermi energy 
for the vacuum that we have introduced in this paper 
is more natural than what Banerjee suggested.


\section{Summary and Discussion}
We have examined the vacuum charge for a fermion field
model based on the one-dimensional Dirac equation with
an external potential. This model is a typical one with 
which fractionization of the vacuum charge has been 
illustrated in the literature. We classified the 
fractional vacuum charges so far examined in the 
literature into two types, A and B. When the external
potential of the model is continuously varied, the 
vacuum charge of type A remains unchanged, whereas that 
of type B varies continuously. We assumed that there 
is no zero mode for the Dirac equation so that the 
charge fractionization of type A does not appear. We 
have focussed on the charge fractionization of type B. 
We pointed out that the topology that has been 
mentioned in the literature in connection with type B 
is of a trivial nature.

The vacuum charge, which is the difference between the 
charge of the perturbed vacuum and that of the free vacuum, 
is in the form of the difference between two divergent 
series. The result depends on how the difference is 
calculated. We proposed a natural way of resolving this 
ambiguity. For the negative energy states of the free
vacuum, we assume a Fermi energy $E_F <0$. All the 
energy levels between 0 and $E_F$ are occupied and those
below $E_F$ are empty. When the external potential is 
switched on the Fermi energy shifts, $E_F \to E'_F$ but 
the one-to-one correspondence between the $E$s and $E'$s 
is maintained. We eventually let $E_F \to -\infty$ and 
$E'_F \to -\infty$. This device is instrumental in 
clearly keeping track of the number of the particles 
in the vacuum. 

In Sec. III we examined a model that extends to the
entire one-dimensional space. We divided the total
charge of the vacuum $ Q$ into two parts, $ Q_c $ and 
$ Q_d $. The $ Q_c $ is the part of the charge that has 
been examined extensively in the literature. In order 
to see the charge conservation $ Q = Q_c + Q_d =0$, it 
is crucial to include $ Q_d $ that is related to the 
change in the Fermi energy due to the external potential. 
Each of $ Q_c$ and $ Q_d$ may take a fractional value 
but we pointed out that this is a mathematical artifact 
that arises from replacing $\sum_E$ with $(R/\pi)\int dk$. 
In this continuum limit the discrete energy levels 
become effectively smeared out. We emphasized that the
level spacing and the energy shift in this situation are 
of the same order of magnitude. The number of smeared 
out energy levels in a specified range becomes fuzzy. 
This artifact, however, does not jeopardize the 
conservation of the total charge.

The model that we examined in Sec. III can be taken as
a one-dimensional simulation of the relativistic shell
model of the atomic nucleus. If we take the fractional 
charge literally, this leads to bizarre consequences as 
we discussed towards the end of Sec. III. If the shell 
model potential is constructed in a self-consistent
manner, we expect that the diffuse distribution 
$\rho_d (x)$ becomes confined to a finite region 
around the nucleus. The total nuclear charge is not
fractionalized.

In Sec. IV we examined the vacuum charge of a finite bag
confined within $|x|<R$. Since $R$ is finite and not very 
large, we do not replace $\sum_E$ with $\int dk$. 
Nevertheless a fractional charge can appear when we use 
the summation method, Eq. (\ref{49}), that is related to 
the split-point regularization of the current operator. 
This occurs in the MIT bag model as well as in the chiral 
bag model. We concur with Banerjee \cite{23} in arguing 
that this outcome of Eq. (\ref{49}) is unnatural. From
our point of view, no fractionization of the charge or 
the quark number appears in the bag models.
   
What we have discussed in this paper may have relevance
to the charge renormalization of quantum electrodynamics
(QED). Suppose a test charge $e_0$ is placed in the 
vacuum. The $e_0$ is renormalized to $e$,
\begin{equation}
e = e_0 (Z_2 /Z_1)\sqrt{Z_3} .
\label{57}
\end{equation}
The Ward identity leads to $Z_1 =Z_2$. We are still left
with $Z_3$ which is due to the vacuum polarization; 
see, e.g. Ref. \cite{24}. If we calculate $Z_3$ by 
perturbation theory, it is represented by a divergent 
integral. If we cut-off the integral, $Z_3$ depends on 
the cut-off parameter. Depending on the choice of the 
cut-off parameter, $Z_3$ can take any fractional value. 
This was probably the first example in which a fractional 
charge appeared. The process that is responsible for the 
charge renormalization is typically through the 
electron-positron pair creation in the virtual state. 
Because the charge of this pair is zero, however, it 
cannot change the charge of the system unless part of 
the charge somehow disappears. The mechanism of 
$Z_3 \neq 1$ is not clear. If we consider the same 
problem but assuming that the space is confined in a 
finite cavity, the situation will be similar to the 
one that we discussed in Sec. IV. Whatever happens to
the system, the total charge is confined within the 
cavity and hence it is conserved. This seems to mean 
$Z_3 =1$. If we let the radius of the cavity tend to 
infinity, we will still have $Z_3 =1$. Would it be 
possible to reformulate QED such that we do not have 
to consider the charge renormalization? \\

This work was supported by the Natural Sciences and 
Engineering Research Council of Canada. 

\nopagebreak


\end{document}